\documentclass[12pt]{article}
\advance \textheight by \topskip
\makeatletter
\@addtoreset{equation}{section}
 
\makeatother

\newcommand{\be}{\begin{equation}}
\newcommand{\ee}{\end{equation}}
\newcommand{\bea}{\begin{eqnarray}}
\newcommand{\eea}{\end{eqnarray}}
\newcommand{\dd}{\partial}

\def\>{\rangle}
\def\<{\langle}

\begin{document}

\title{
{\bf Dynamical Evolution of the Extra Dimension in Brane Cosmology}}


\author{
{\sf   N. Mohammedi} \thanks{e-mail:
nouri@celfi.phys.univ-tours.fr}$\,\,$${}$
\thanks{This work was carried out at the Department of Applied Mathematics
and Theoretical Physics, Cambridge, UK.} 
\\
{\small ${}${\it Laboratoire de Math\'ematiques et Physique Th\'eorique,}} \\
{\small {\it Universit\'e Fran\c{c}ois Rabelais,}}\\
{\small {\it Facult\'e des Sciences et Techniques,}}\\
{\small {\it Parc de Grandmont, F-37200 Tours, France.}}}
\date{}
\maketitle
\vskip-1.5cm

\vspace{2truecm}

\begin{abstract}

\noindent
The evolution of the extra dimension is investigated in 
the context of brane world cosmology. New cosmological solutions are 
found. In particular, solutions in the form of waves travelling along the extra dimension are
identified.

\end{abstract}

\newpage

%

\setcounter{equation}{0}

\section{Introduction}

The brane world scenario \cite{sundrum1,sundrum2} 
stipulates that our four-dimensional Universe
(the brane)
is embedded in a higher dimensional space-time (the bulk). This
approach differs from the usual Kaluza-Klein ideas in  that the
size 
of the extra dimensions could be large. The concept of large extra
dimensions
might have phenomenological consequences in particle 
physics \cite{arkani1,arkani2,arkani3}. In
particular,
it could lead to a solution of the hierarchy problem (the 
problem of why the electroweak energy scale and the Planck
energy scale  are far apart from one
another). 
\par
Another important ingredient of the brane world scenario is that 
matter is confined to the brane and the only communication between
the brane and the bulk is through gravitational interaction (or some
other
dilatonic matter). Newton's law of gravitational interaction, 
as we know it in our Universe, would then arise 
as a very good approximation in this context. The brane world picture
relies on a $Z_2$ symmetry and is inspired from string theory and its 
extensions \cite{horava}.
\par
There are, however, numerous questions that one would like to address
in the brane world scenario. The first, and most difficult, question
regards the mechanism by which matter is forced to live on the brane
only. In the absence of any answers to this question, one would 
accept it as a hypothesis and looks for experimental evidence for this
scenario.
The natural laboratory for testing the ideas of this new theory (and
of string theory in general, see \cite{quevedo,easson,copeland} for reviews)
would be in cosmology. It is therefore essential to understand the
theoretical 
implications of a brane world based cosmology \cite{binetruy}
(see \cite{langlois,brax} for some pedagogical reviews). 
\par
The most stricking feature of brane world cosmology 
\cite{lot1,lot2,lot3,lot4,lot5,lot6,lot7,lot8,lot9}
is the fact that
the square of the Hubble parameter on the brane is proportional to the square of
the energy density of the brane. This is in contrast to 
the situation, described by the Friedmann equation, in the standard four-dimensional 
cosmology. 
This  proportionality between 
the Hubble parameter and the energy density is due primarily to 
the requirement that the incuded metric on the brane
is  that of Friedmann-Robertson-Walker (FRW).
\par
Another important point in the brane world scenario regards the size
of the extra dimension. As mentioned previously, this size can be 
arbitrarily large. It is therefore crucial to have an idea on how 
this extra dimension evolves. This problem can be naturally examined 
in the context of brane world cosmology. In most cosmological studies
of the brane world, the extra dimension is taken to be constant.
In this note, we will allow the size of the extra dimension to 
evolve dynamically. Some interesting solutions are consequently found. 
\par
We start, in section 2,  by a brief review of the equations of motion 
behind the brane world cosmology.
The assumptions used in this scenario are also revisited.
It is shown that there is a
great deal of freedom in choosing the bulk metric.
In section 4, we make an ansatz for the form of the size 
of the extra dimension and determine some new cosmological 
solutions.

\section{ A review of brane cosmology}

The model studied here is that of a single  brane (our Universe) embedded in a 
five-dimensional spacetime (the bulk) whose coordinates are 
$x^\mu=\left(t,r,\theta,\phi,y\right)$ with $\mu,\nu,\dots=0,\dots ,4$.
The brane is located at $y=0$.
Our starting point is the five-dimensional Einstein's equations 
\bea
{\cal E}_{\mu\nu}\equiv\alpha \left(R_{\mu\nu}-{1\over 2}Rg_{\mu\nu}\right)
 -{\Lambda\over 2}  g_{\mu\nu}- T_{\mu\nu}=0 \,\,\,\,.
\eea
Here $\alpha$ is the gravitational coupling constant and $\Lambda$ is
a possible cosmological constant. The five-dimensional 
metric is specified by the 
line element \cite{binetruy}
\be
ds^2=-A(t,y)^2dt^2 +B(t,y)^2\left[{dr^2\over 1-kr^2}
+r^2\left(d\theta^2+\sin\left(\theta\right)^2d\phi^2\right)\right]
+C(t,y)^2dy^2\,\,\,\,\,.
\label{metric}
\ee
The functions $A(t,y)$, $B(t,y)$ and $C(t,y)$ depend on the 
variable $y$ through its modulus $|y|$ only. 
This is in order to realise the $Z_2$ symmetry; a crucial point in 
the brane world scenario. The parameter $k$ ($k=-1,0,1$) is the 
spacial curvature of a maximally symmetric three-dimensional metric.
To complete the cosmological
setting, the energy-momentum tensor $T_{\mu\nu}$ is taken to have the form
\be
T^\mu_\nu={1\over C}\, {\rm diag}\left[-\rho(t),p(t),p(t),p(t),0\right]\,
\delta\left(y\right)\,\,\,\,.
\ee
This choice is compatible with the metric and the matter is indeed located
on the brane only. It is clear that our description breaks down whenever
$C(t,y)=0$. 
We require, also, the matter on the brane to obey 
the equation of state
\be
p=\omega \rho\,\,\,\,\,,
\label{eqstate}
\ee
where $\omega$ is a constant. 
\par
The equations of motion lead to four equations: ${\cal E}_{00}$, 
${\cal E}_{11}={\cal E}_{22}={\cal E}_{33}$, ${\cal E}_{44}$
and  ${\cal E}_{04}$. These are, respectively, given by\footnote{
{\underline{Notation}}: If $f(|y|)$ and $h(|y|)$ are two functions, then  ${df\over dy}=f'{d|y|\over
dy}= f'\left[2\Theta\left(y\right)-1\right]$,
where $\Theta\left(y\right)$ is the Heaviside function and $f'$
denotes the derivative of $f$ with respect to its argument
$|y|$. Consequently, $\left({df\over dy}\right)\left({dh\over dy}\right)
=f'h'$ and 
${d^2f\over dy^2}=f''+2f'\delta\left(y\right)$, where $f''$ is the
second derivative of $f$ with respect to $|y|$.
We also use a dot to denote a derivative with respect to the time
coordinate $t$.} 
\bea
{1\over C}\,\rho\,\delta(y) &=&
3\alpha\left\{-{1\over BC^2}\left[B''+2B'\delta(y)\right]
+{B'\over BC^2}\left({C'\over C}-{B'\over B}\right)
+{\dot{B}\over BA^2}\left({\dot{B}\over B}+{\dot{C}\over C}\right)
+{k\over B^2}\right\} \nonumber \\
& + & {\Lambda\over 2} \,\,\,\,\,\,,
\nonumber\\
{1\over C}\,p\,\delta(y) &=&
\alpha\left\{2{1\over BC^2}\left[B''+2B'\delta(y)\right]
+{1\over AC^2}\left[A''+2A'\delta(y)\right]
+{B'\over BC^2}\left({B'\over B}-2{C'\over C}\right)\right.
\nonumber \\
&+&{A'\over AC^2}\left(2{B'\over B}-{C'\over C}\right)
-{\dot{B}\over BA^2}\left({\dot{B}\over B}+2{\dot{C}\over C}\right)
+{\dot{A}\over A^3}\left({\dot{C}\over C}+2{\dot{B}\over B}\right)
\nonumber\\
&-& 2{\ddot{B}\over BA^2}-{\ddot{C}\over CA^2}
-{k\over B^2}\left.\right\} -{\Lambda\over 2}\,\,\,\,\,\,,
\nonumber \\
0 &=&
3\alpha\left\{
{B'\over B}\left({B'\over B}+{A'\over A}\right)
+{C^2\dot{B}\over BA^2}\left({\dot{A}\over A}-{\dot{B}\over B}\right)
-{C^2\ddot{B}\over BA^2}
-{kC^2\over B^2}\right\} -{\Lambda\over 2}C^2 \,\,\,\,\,\,,
\nonumber \\
0&=& 3\alpha\left\{
{\dot{B}A'\over BA} +{\dot{C}B'\over CB}-{\dot{B}'\over B}
\right\}\left[2\Theta\left(y\right)-1\right]
\,\,\,\,\,.
\label{eqmotion}
\eea
Matching the delta functions on both sides of the first two equations
leads to
\bea
\rho &=&- 6\alpha {B'_0\over B_0 C_0}\nonumber \\
p &=&-{2\over 3}\rho + 2\alpha {A'_0\over A_0 C_0} \,\,\,\,.
\label{match}
\eea
Here the subscript $0$ means that the functions are evaluated at $y=0$
(that is $A_0=A(t,0)$, and so on). Once this matching is carried out,
the delta function contributions cancel out and the equations become
valid everywhere. Notice also that the obtained  equation of state is not of the 
form $p=\omega\rho$ but a time dependent one. 
\par
It is remarkable that the first three equations in (\ref{eqmotion}) can be expressed
in terms of the  single function \cite{binetruy}
\be
F(t,y)={\left(BB'\right)^2\over C^2}-{\left(B\dot{B}\right)^2\over
A^2} -kB^2 \,\,\,\,.
\ee
Assuming that ${\cal{E}}_{04}$ (that is, the last  equation in
(\ref{eqmotion}) )
is satisfied, then the components
${\cal E}_{00}$ and ${\cal E}_{44}$ of the equations of motion can be
cast, respectively,  in the form
\bea
F'&=& {\Lambda\over 3\alpha}B^3 B' \nonumber \\
\dot{F} &=& {\Lambda\over 3\alpha}B^3 \dot{B}\,\,\,\,.
\label{Feq}
\eea
On the other hand, the component ${\cal{E}}_{11}$ of the equations
of motion (with the help of  ${\cal{E}}_{00}$,  ${\cal{E}}_{44}$
and ${\cal{E}}_{04}$) can be written as
\be
{\dd\over \dd t}\left({F'\over B'}\right)= {\Lambda\over \alpha}
\dot{B}B^2\,\,\,\,\,.
\ee
This last equation is identically satisfied due to (\ref{Feq}).
\par
Therefore, by integration of (\ref{Feq}), one obtains the first integral of motion
\be
{\left(BB'\right)^2\over C^2}-{\left(B\dot{B}\right)^2\over
A^2} -kB^2={\Lambda\over 12\alpha}B^4 +{\cal C}\,\,\,\,\,,
\label{firstint}
\ee
where ${\cal C}$ is a constant of integration. 
This last  equation can be used to determine the unknown function
$A$. 
Indeed, this is  given by  
\be
A^2=\dot{B}^2 
\left[
{\left(B'\right)^2\over C^2}-k -{\Lambda\over 12\alpha}B^2
-{{\cal C}\over B^2}\right]^{-1}\,\,\,\,\,.
\label{A}
\ee
Therefore $A$ is entirely expressed in terms of $B$, $C$ and their derivatives.
Substituting for A in the last equation of  (\ref{eqmotion}), yields the differential
equation
\be
\dot{B}\,{\dd\over \dd|y|}\left\{\ln\left[
{\left(B'\right)^2\over C^2}-k -{\Lambda\over 12\alpha}B^2
-{{\cal C}\over B^2}\right]\right\}=2B'{\dot{C}\over C}\,\,\,\,\,.
\label{onlyeq}
\ee
This is the only equation that the two unknown functions $B$ and $C$
have to satisfy. Hence, it has many solutions in general. Furthermore,
one can use it to exclusively fixe the $y$ dependence of $B$ and $C$ (and
consequently $A$) but not their time dependence.

\section{The brane equations}

The most crucial relation in brane world cosmology is 
equation (\ref{A}).
It leads, when evaluated at $y=0$ together with the use of the matching 
conditions (\ref{match}), to the expression
\be
A_0^2= {\dot{B}_0^2\over B_0^2} \left\{{\rho^2\over 36\alpha^2}
-{k\over B_0^2}-{\Lambda\over 12\alpha}-{{\cal{C}}\over B_0^4}\right\}^{-1}
\,\,\,\,\,.
\label{A0=1}
\ee
This is the time dependent value of $A_0$ as determined by the equations of motion.  
If one imposes the  temporal gauge $A_0=1$, then one obtains the
Friedmann like relation \cite{binetruy}
 \be
H^2\equiv {\dot{B}_0^2\over B_0^2}={\rho^2\over 36\alpha^2}
-{k\over B_0^2}-{\Lambda\over 12\alpha}-{{\cal{C}}\over B_0^4}
\,\,\,\,\,.
\label{binetal}
\ee
The unusual proportionality between $H^2$ and $\rho^2$, together with the 
presence of the term involving ${\cal{C}}$,  are the main caracteristics of brane world
cosmology \cite{binetruy}. This equation, however, is obtained only when one supposes
that $A_0=1$. This assumption is equivalent to demanding that the
metric on the brane is a FRW metric
at all times. It would be, therefore, desirable to consider other gauges
and to determine their cosmological implications. However, this is 
not the issue of this note.
\par
Let us first examine the time dependence
of the two functions $B$ and $C$. We start by exploring the consequences of imposing
the equation of state (\ref{eqstate}). Notice first that  using the
${\cal E}_{04}$ 
component of the 
equations of motion at $y=0$, leads to the conservation equation
\be
B_0{\dot \rho}+3\dot{B}_0\left(\rho + p\right)=0\,\,\,\,\,.
\label{conserv}
\ee
It is, therefore, natural to interpret $B_0$ as the scale factor of
our
Universe.
Moreover, if the equation of  state $p=\omega\rho$ holds then  one obtains
for the energy density 
\be
\rho=\beta B_0^{-3(1+\omega)}\,\,\,\,,
\label{rho2}
\ee
where $\beta$ is an integration constant. On the other hand, 
for the matching equations (\ref{match}) to yield the equation of state
$p=\omega\rho$, we must have
\bea
 2\alpha {A'_0\over A_0 C_0} &=& \gamma \rho 
\,\,\,\,\,,
\label{rho1}
\eea
where $\gamma$ is a constant and we have the identification
\be
\omega=-{2\over 3} + \gamma\,\,\,\,\,.
\ee
In order for the  energy density $\rho$ to decrease when $B_0$ grows,
one must have $\gamma >-1/3$. We are of course assuming that the
Universe is expanding.
\par
Using the component ${\cal E}_{04}$ of the equations of motion and 
the expression of $\rho$ in (\ref{match}), equation (\ref{rho1}) can be cast in the form
\be
{\dd \over \dd t}\left[\ln\left({B'_0B_0^{3\gamma}\over C_0}
\right)\right]=0\,\,\,\,.
\ee
Therefore
\be
C_0=\lambda B'_0B_0^{3\gamma}\,\,\,\,\,,
\label{C0}
\ee
where $\lambda$ is a constant.
Furthermore,
by comparing the two expressions of $\rho$ in (\ref{rho2}) and (\ref{match}), we deduce
that
\bea
\beta B_0^{-(1+3\gamma)} &=& -6\alpha{B'_0\over B_0 C_0}\,\,\,\,\,. 
\eea
Replacing for $C_0$ in the last equation fixes the constant $\lambda$
to
\be
\lambda=-{6\alpha\over\beta}\,\,\,\,.
\ee
The equation of state is, therefore, given by $p=\omega \rho$ provided
that the function $C$ obeys the  brane condition 
$C_0=\lambda B'_0B_0^{3\gamma}$.

\section{Solutions with a dynamical fifth dimension}

One of the simplest cases studied so far corresponds
to setting  $C(t,y)=1$ at all times \cite{binetruy}. 
We would like to present in this section another solution 
in which the fifth dimension is evolving. This is in the spirit
of dynamical compactification in Kaluza-Klein theories \cite{kaluza}. 
Let us assume that $C$ is a function of $B$ only. Namely,
\be
C(t,y)=C(B(t,y))\,\,\,\,.
\label{C(B)}
\ee
This leads immediately to the relation
\be
2B'{\dot{C}\over C}=\dot{B}{\dd \over \dd|y|}\left(\ln
C^2\right)\,\,\,\,\,.
\ee
Therefore, equation (\ref{onlyeq}) can be integrated and one  obtains
the first order differential equation 
\be
\left(B'\right)^2= \xi C^4 +C^2\left[k +{\Lambda\over 12\alpha}B^2
+{{\cal C}\over B^2}\right]\,\,\,\,\,,
\label{diffeq}
\ee
where $\xi(t)$ is a  function of integration
(it will be shown later that $\xi$ is a constant). This equation
determines $B$ once the function $C(B)$ is known.
Its right hand side is a functional of $B$ only and therefore
can be formally integrated as
\bea
 E\left(B\right)&\equiv&\int dB\left\{
\xi C^4 +C^2\left[k +{\Lambda\over 12\alpha}B^2
+{{\cal C}\over B^2}\right]\right\}^{-1/2}
\nonumber\\
E\left(B\right)&=&\pm |y| + \zeta\left(t\right)\,\,\,\,,
\label{E}
\eea
where $\zeta(t)$ is a function of integration. The last equation 
determines $B$ in terms of $|y|$, $\xi(t)$ and $\zeta(t)$.
\par
Let us now impose the constraint
$C_0= \lambda B'_0B_0^{3\gamma}$. Replacing for $B'_0$ using (\ref{diffeq}), one 
find the following expression for $C_0^2$
\be
C_0^2=
{{1\over  \xi}\left\{{1\over \lambda^2B_0^{6\gamma}}
-\left[k +{\Lambda\over 12\alpha}B_0^2
+{{\cal C}\over B_0^2}\right]\right\}}\,\,\,\,\,.
\label{C0(B)}
\ee
In order for this expression of $C_0$ to be in accordance with our
assumption in (\ref{C(B)}), the function $\xi(t)$ must be constant in
time.
Since $C_0(B_0)$ is obtained  from $C(B)$ by simply replacing $B$ by
$B_0$, we deduce that 
\be
C^2=
{{1\over  \xi}\left\{{1\over \lambda^2B^{6\gamma}}
-\left[k +{\Lambda\over 12\alpha}B^2
+{{\cal C}\over B^2}\right]\right\}}\,\,\,\,\,.
\label{choice}
\ee
Finally,  
we should mention that the function $A$ in (\ref{A}) is, in this case,  given by
\be
A^2={\dot{B}^2 \over {\xi} C^2}\,\,\,\,\,.
\label{AC(B)}
\ee
The $y$ dependence of $B$ is therefore determined by evaluating the integral
\be
E\left(B\right)=\sqrt{\xi}\lambda^2\int dB{B^{(1+6\gamma)}\over
\sqrt{B^2-\lambda^2kB^{(2+6\gamma)}-{\lambda^2\Lambda\over 12\alpha}
B^{(4+6\gamma)} -\lambda^2{\cal{C}}B^{6\gamma}}}\,\,\,\,\,.
\ee
It is clear that this integral depends on the the values of $\gamma$
and the other parameters $k$, $\Lambda$ and ${\cal{C}}$.
\par
As an illustration, we will consider here the case corresponding to
$\gamma=0$ and $\Lambda\ne 0$. This means that the matter on the brane obeys the
equation
of state $p=-2\rho/3$ and is therefore not an ordinary matter.
The integral can be calculated and we find, using (\ref{E}),
the following expression for $B$
\bea
B^2 &=&
2\sqrt{
{3\alpha{\cal{C}}\over \Lambda}
-\left({3\alpha\over\lambda^2 \Lambda}\right)^2
\left(1-\lambda^2 k\right)^2}\,
\sinh\left\{\sqrt{-{\Lambda\over 3\alpha\xi\lambda^2}}
\left[\pm |y| + \zeta - a\right]\right\}
\nonumber\\
&+& {6\alpha\over \lambda^2\Lambda}\left(1-\lambda^2 k\right)
\,\,\,\,\,,
\label{gamma=0}
\eea
where $a$ is a constant of integration.
\par
The case $\gamma=0$ and $\Lambda=0$ is also worth studying. In this
case, we find the following expression for $B^2$
\be
B^2={1\over\left(1-\lambda^2 k\right)}\left[
\lambda^2{\cal{C}}+
{\left(1-\lambda^2 k\right)\over \xi\lambda^4}
\left(\pm|y|+\zeta-b\right)^2\right]\,\,\,\,,
\label{gamma=00}
\ee
where $b$ is a constant of integration. Here also we are considering
the case where $\left(1-\lambda^2 k\right)$ does not vanish.
\vskip1.5cm
\par
\noindent
{{\it{Cosmological solutions}}} :
\vskip1.5cm
\par
\noindent
The time dependence of $B$, for $\gamma=0$ and $\Lambda\ne 0$, is determined by solving
the constaint $A_0^2=1$. Equation (\ref{AC(B)}) yields in this case
\be
B_0^2{\dot{B}}_0^2=
\left({1\over \lambda^2}-k\right)B_0^2
-{\Lambda\over 12\alpha}B_0^4-{\cal{C}}
\,\,\,\,\,\,.
\ee
We will assume that $\left({1\over \lambda^2}-k\right)$ is different
from zero. The solution to the above equation is given by
\be
B_0^2={6\alpha\over\Lambda}\left({1\over \lambda^2}-k\right)
+a_1\exp\left(\sqrt{-{\Lambda\over 3\alpha}}t\right)
+a_2\exp\left(-\sqrt{-{\Lambda\over 3\alpha}}t\right)
\,\,\,\,,
\ee
where $a_1$ and $a_2$  are two constants related by
\be
a_1a_2=\left({3\alpha\over \Lambda}\right)^2
\left({1\over \lambda^2}-k\right)^2-{3\alpha{\cal{C}}\over \Lambda}
\,\,\,\,.
\ee
The expression found for $B_0^2$ must agree with that 
obtained from (\ref{gamma=0}) upon setting $y=0$. This requirement 
leads to 
\bea
2\sqrt{-a_1a_2}\,\,\sinh\left[\sqrt{-{\Lambda\over 3\alpha\xi\lambda^2}}
\left(\zeta-a\right)\right]
=a_1\exp\left(\sqrt{-{\Lambda\over 3\alpha}}t\right)
+a_2\exp\left(-\sqrt{-{\Lambda\over 3\alpha}}t\right)
\,\,\,\,\,.
\eea
This last equation allows for the determination of the unknown
function $\zeta(t)$.  
\par
It is clear that $\zeta$ depends on the choice
one makes for the two constants $a_1$ and $a_2$. An interesting
case corresponds to taking $a_2=-a_1$ and such that
\be
a_1^2=
{3\alpha{\cal{C}}\over\Lambda}-\left({3\alpha\over \Lambda}\right)^2
\left({1\over \lambda^2}-k\right)^2 \,\,\,\,\,.
\ee
The function $\zeta(t)$ is in this case linear and is given by
\be
\zeta\left(t\right)=\lambda\sqrt{\xi}t+a
\,\,\,\,\,.
\ee
The importance of this special case stems from the fact that
$B^2\left(t,y\right)$ is a function of the combination
$\left(\pm|y|+\lambda\sqrt{\xi}t\right)$. Therefore,
it has the form of a wave travelling in the $y$ direction.
\par
Similarly, the case corresponding to $\gamma=0$ and $\Lambda=0$
leads to 
\be
B_0^2=\left({1\over \lambda^2}-k\right)t^2 +b_1t+b_2\,\,\,\,,
\ee
where the two constants $b_1$ and $b_2$ are related by
\be
{1\over 4}b_1^2=\left({1\over \lambda^2}-k\right)b_2-{\cal{C}}
\,\,\,\,.
\ee
Finally, the comparison of this last expression with that
obtained from   (\ref{gamma=00}) upon setting $y=0$,
yields
\be
\zeta\left(t\right)=\pm\lambda\sqrt{\xi}
\left(t+{b_1\over 2}{\lambda^2\over\left(1-\lambda^2 k\right)}\right)
+b\,\,\,\,\,.
\ee
This shows also that $B^2\left(t,y\right)$ is a wave travelling in the $y$ direction.

In conclusion, we have in this note revisited the scenario of brane world 
cosmology. We have found a class of solutions in which the fifth dimension
evolves dynamically. It is interesting to notice that these solutions 
are in the form of waves travelling in the $y$ direction.
This has been shown for some special values of the parameters $\gamma$ and
$\Lambda$. However, by a change of the time coordinate, one can show that
these waves are always present. Indeed, let the new time coordinate be
$\tau=\zeta(t)$, then the metric corresponding to our solution 
is given by
\be
ds^2=-{1\over \xi C^2}\left({dB\over d\tau}\right)^2
d\tau^2 +B^2\left[{dr^2\over 1-kr^2}
+r^2\left(d\theta^2+\sin\left(\theta\right)^2d\phi^2\right)\right]
+C^2dy^2\,\,\,\,\,,
\ee
where $C^2$, as a functional of $B$, is given by the expression 
written in (\ref{choice}). On the other hand,
$B$ is a function of the variable $\pm|y|+\tau$ only. This  can be seen
from the second equation of (\ref{E}).
\par
If we now demand that $\tau$ is the cosmological time and that
at $y=0$ (on the brane) the metric is a FRW metric, then the
time evolution of $B_0$ is determined by solving the equation
\be
\left({dB_0\over d\tau}\right)^2=\xi C^2_0\,\,\,\,\,.
\ee
This is precisely equation (\ref{binetal}) where $t$ is
simply replaced by $\tau$.

\vskip1.0cm
\noindent
{\bf {Acknowledgements}} : I would like to thank A. C. Davis and
N. Grandi for discussions. I would like also to extend my thanks 
to all members of D.A.M.T.P at the University of
Cambridge for their kind hospitality.
This research is supported by the CNRS.

\end{document}